\documentclass[openacc]{rstransa}

\usepackage{amsmath}
\usepackage{amssymb}
\usepackage{graphicx}
\usepackage[normalem]{ulem}
\usepackage[dvipsnames]{xcolor}

\usepackage[numbers,sort&compress]{natbib}

\usepackage{hyperref}
\usepackage{enumitem}

\newcommand*\aap{A\&A}

\newcommand*\apj{ApJ}
\newcommand*\apjl{ApJ}

\newcommand*\apjs{ApJS}

\newcommand*\mnras{MNRAS}

\newcommand*\nat{Nature}

\newcommand*\procspie{Proc.~SPIE}

\newcommand*\skytel{S\&T}
\newcommand*\solphys{Sol.~Phys.}

\newcommand*\ssr{Space~Sci.~Rev.}

\newcommand*\ptrsa{Phil. Trans. R. Soc. A.}

\begin{document}

\title{High-resolution wave dynamics in the lower solar atmosphere}

\author{
D.~B.~Jess$^{1,2}$, P.~H.~Keys$^{1}$, M.~Stangalini$^{3}$ and S.~Jafarzadeh$^{4,5}$
}

\address{
$^{1}$Astrophysics Research Centre, School of Mathematics and Physics, Queen's University Belfast, Belfast, BT7 1NN, UK\\
$^{2}$Department of Physics and Astronomy, California State University Northridge, Northridge, CA 91330, USA\\
$^{3}$ASI Italian Space Agency, Via del Politecnico snc, I-00133 Rome, Italy\\
$^{4}$Rosseland Centre for Solar Physics, University of Oslo, P.O. Box 1029 Blindern, NO-0315 Oslo, Norway\\
$^{5}$Institute of Theoretical Astrophysics, University of Oslo, P.O. Box 1029 Blindern, NO-0315 Oslo, Norway\\
}

\subject{astrophysics, observational astronomy, solar system, spectroscopy, wave motion}


\corres{David B. Jess\\
\email{d.jess@qub.ac.uk}}

\begin{abstract}
The magnetic and convective nature of the Sun's photosphere provides a unique platform from which generated waves can be modelled, observed, and interpreted across a wide breadth of spatial and temporal scales. As oscillations are generated in-situ or emerge through the photospheric layers, the interplay between the rapidly evolving densities, temperatures, and magnetic field strengths provides dynamic evolution of the embedded wave modes as they propagate into the tenuous solar chromosphere. A focused science team was assembled to discuss the current challenges faced in wave studies in the lower solar atmosphere, including those related to spectropolarimetry and radiative transfer in the optically thick regions. Following the Theo Murphy international scientific meeting held at Chicheley Hall during February 2020, the scientific team worked collaboratively to produce 15 independent publications for the current Special Issue, which are introduced here. Implications from the current research efforts are discussed in terms of upcoming next-generation observing and high performance computing facilities.
\end{abstract}

\maketitle



\begin{figure}[!t]
\centering\includegraphics[width=0.344\textwidth]{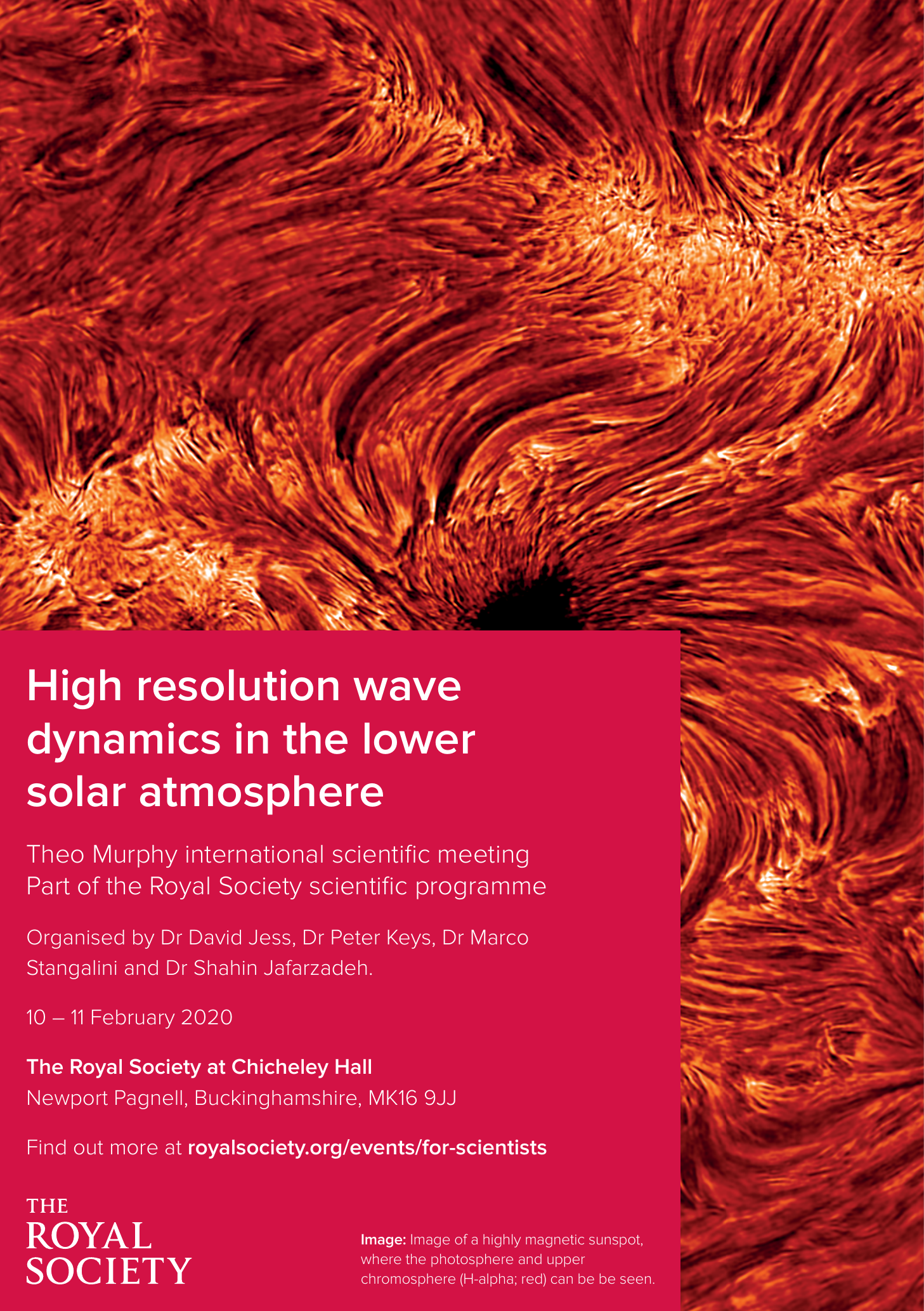}
\centering\includegraphics[width=0.635\textwidth]{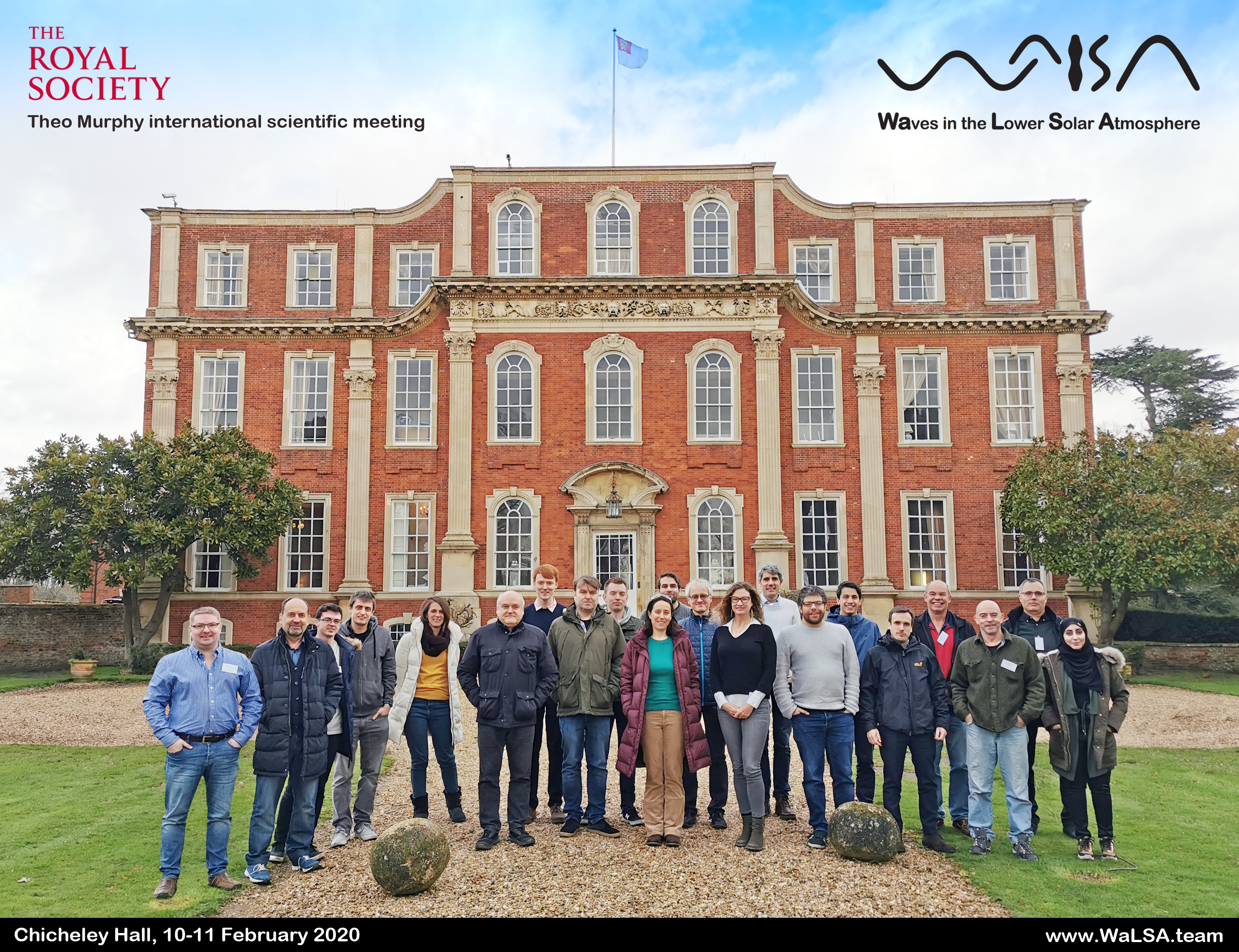}
\caption{The Royal Society advertisement for the Theo Murphy Meeting {\it{``High resolution wave dynamics in the lower solar atmosphere''}} (left), alongside a picture of the assembled team in front of Chicheley Hall during the meeting on ${10-11}$~February~2020 (right). From left-to-right, the people visible in the photo are Dr.~David Jess, Dr.~Robert Sych, Mr.~Samuel Skirvin, Dr.~Peter Keys, Dr.~Elena Khomenko, Prof.~Tony Arber, Mr.~Conor MacBride, Dr.~Gary Verth, Dr.~Samuel Grant, Dr.~Rebecca Centeno Elliott, Dr.~Ben Snow, Dr.~Matthias Rempel, Dr.~Aimee Norton, Dr.~Alex Feller, Mr.~Callum Boocock, Mr.~Daniele Calchetti, Dr.~Marco Stangalini, Dr.~Bernhard Fleck, Prof.~Stuart Jefferies, Dr.~Shahin Jafarzadeh, and Ms.~Anwar Aldhafeeri.}
\label{fig:RS_flyer_team}
\end{figure}

\section{Introduction}
\label{sec:introduction}
The topic of waves and oscillations in the Sun's atmosphere has been a focal point of solar physics research since their first detection over half a century ago \citep{1962ApJ...135..474L, 1963ApJ...138..631N}. The concentrated magnetic fields that permeate the entire solar surface are able to act as efficient waveguides for the oscillatory behaviour, hence providing conduits for the flow of wave energy upwards into the outer regions of the solar corona \citep{2015SSRv..190..103J}. With the lower solar atmosphere, comprising of the photosphere and chromosphere, being predominantly visible in the infrared, optical, and ultraviolet portions of the electromagnetic spectrum, it becomes possible to observe these tenuous regions using a combination of ground-based and space-borne facilities. 

In recent decades, modern upgrades to existing observatories \citep[e.g., the Dunn, Swedish, and Goode Solar Telescopes;][]{1969S&T....38..368D, 2003SPIE.4853..341S, 2010AN....331..636C}, in addition to the construction and/or launch of new observing facilities \citep[e.g., Hinode, the Interface Region Imaging Spectrograph, the National Science Foundation's Daniel K. Inouye Solar Telescope;][]{2007SoPh..243....3K, 2014SoPh..289.2733D, 2016AN....337.1064T}, have paved the way for rapid advancements to be made in the field of lower atmospheric dynamics. In particular, the high spatial, temporal, and spectral resolutions now available from the cutting-edge observatories, in combination with polarimetric capabilities,  we have at our disposal have made the studies of small-scale oscillatory phenomena possible, resulting in numerous high-impact publications in recent years \citep[e.g.,][]{2014Sci...346D.315D, 2017Sci...356.1269M, 2017NatSR...743147S, 2018NatPh..14..480G, 2019Sci...366..890S, 2020NatAs...4..220J, 2020NatAs.tmp..149J}.

\begin{table}[!h]
\caption{Summary of contributions to the WaLSA research team and the themed Special Issue of the {\it{Philosophical Transactions of the Royal Society A}}, organised alphabetically by country to document the affiliated research institutes.}
\label{tab:WaLSAinstitutes}
\begin{tabular}{|l|l|}
\hline
Country & Research Institute  \\
\hline
\hline
\scriptsize{Australia} & \scriptsize{School of Mathematical and Physical Sciences, University of Newcastle} \\	
\hline
\scriptsize{Belgium} & \scriptsize{Centre for mathematical Plasma Astrophysics (CmPA), KU Leuven} \\	
\hline
\scriptsize{Chile} & \scriptsize{School of Engineering, Pontificia Universidad Cat\'{o}lica de Chile} \\
\hline
\scriptsize{China} & \scriptsize{National Astronomical Observatories, Chinese Academy of Science} \\
 & \scriptsize{School of Astronomy and Space Sciences, University of Chinese Academy of Sciences} \\	
\hline
\scriptsize{Germany} & \scriptsize{Leibniz-Institut f{\"{u}}r Sonnenphysik (KIS)} \\
 & \scriptsize{Max Planck Institute for Solar System Research, G{\"{o}}ttingen} \\
\hline
\scriptsize{Italy} & \scriptsize{ASI Italian Space Agency} \\	
 & \scriptsize{Department of Physics, University of Rome Tor Vergata} \\
 & \scriptsize{INAF-OAR National Institute for Astrophysics} \\
\hline
\scriptsize{Norway} & \scriptsize{Institute of Theoretical Astrophysics, University of Oslo} \\
 & \scriptsize{Rosseland Centre for Solar Physics, University of Oslo} \\
\hline
\scriptsize{Russia} & \scriptsize{Institute of Solar-Terrestrial Physics SB RAS} \\
\hline
\scriptsize{Saudi Arabia} & \scriptsize{Mathematics Department, Majmaah University} \\	
\hline
\scriptsize{Spain} & \scriptsize{Departamento de Astrof{\'{i}}sica, Universidad de La Laguna} \\	
 & \scriptsize{Instituto de Astrof{\'{i}}sica de Canarias} \\	
\hline
\scriptsize{Switzerland} & \scriptsize{Istituto Ricerche Solari Locarno (IRSOL)} \\	
\hline
\scriptsize{United Kingdom} & \scriptsize{Astrophysics Research Centre, Queen's University Belfast} \\
 & \scriptsize{Centre for Geophysical and Astrophysical Fluid Dynamics, University of Exeter} \\	
 & \scriptsize{Department of Mathematics, Northumbria University}	 \\
 & \scriptsize{Plasma Dynamics Group:} \\
 & \hspace{5mm} \scriptsize{School of Mathematics and Statistics, University of Sheffield} \\
 & \hspace{5mm} \scriptsize{Department of Automatic Control and Systems Engineering, University of Sheffield} \\
 & \scriptsize{University College London, Mullard Space Science Laboratory} \\
\hline
\scriptsize{United States of America} & \scriptsize{Ben T. Zinn Combustion Laboratory, Georgia State University} \\
 & \scriptsize{Center for Solar-Terrestrial Research, New Jersey Institute of Technology} \\
 & \scriptsize{College of Science, George Mason University} \\
 & \scriptsize{Department of Physics and Astronomy, California State University Northridge} \\
 & \scriptsize{Department of Physics and Astronomy, Georgia State University} \\	
 & \scriptsize{ESA Science and Operations Department, c/o NASA/GSFC Code 671} \\
 & \scriptsize{Hansen Experimental Physics Laboratory, Stanford University} \\
 & \scriptsize{High Altitude Observatory, NCAR} \\
 & \scriptsize{Institute for Astronomy, University of Hawaii} \\
 & \scriptsize{National Solar Observatory (NSO)} \\
 & \scriptsize{Natural and Applied Sciences, University of Wisconsin} \\
 & \scriptsize{Space Vehicles Directorate, Air Force Research Laboratory} \\
\hline
\end{tabular}
\vspace*{-4pt}
\end{table}

In response to the rapid progress made in the understanding of wave physics in the lower solar atmosphere, a dedicated team was formed in 2017 to continue to build upon developing worldwide expertise and to define new scientific challenges that could be tackled in the years and decades to come. The Waves in the Lower Solar Atmosphere\footnote{WaLSA team initiatives can be viewed at \href{www.WaLSA.team}{www.WaLSA.team}.} (WaLSA) team was formed, and initially consisted of 12 core members spanning 7 countries. Following initial face-to-face meetings in Oslo, Norway, which were funded by the Research Council of Norway (through its Centres of Excellence scheme, the Rosseland Centre for Solar Physics), seed projects based on existing data were created to bolster collaborative partnerships between the team members and their affiliated research institutes. Success of these initial collaborations \citep[e.g.,][]{2018ApJ...869..110S, 2019MNRAS.488L..53K, 2020ApJ...892...49H, 2020A&A...637A..97S, 2020NatAs...4..220J} prompted the WaLSA team coordinators to apply for additional funding to support a dedicated, multi-day meeting where both early career and well established scientists could be invited to join the WaLSA team in order to define new, long-term goals under the scientific umbrella of wave activity in the lower solar atmosphere. 

Funding was provided by The Royal Society Hooke Committee, through the award of a Theo Murphy Meeting that would take place at Chicheley Hall, England, UK, during ${10-11}$~February~2020 (Figure~{\ref{fig:RS_flyer_team}}). Including the 12 core WaLSA team members, the Theo Murphy Meeting welcomed 28 participants from around the world for an intensive 2-day meeting. Following the successful presentations and discussions at the {\it{``High resolution wave dynamics in the lower solar atmosphere''}} meeting, sufficient scientific momentum was generated to enable a special themed issue of the {\it{Philosophical Transactions of the Royal Society A}} (Philos.~Trans.~R.~Soc.~A; \href{https://royalsocietypublishing.org/journal/rsta}{https://royalsocietypublishing.org/journal/rsta}) to be published based on the scientific achievements of the participating cohort and their extended collaborative networks. 

\begin{figure}[!t]
\centering\includegraphics[angle=270, trim=2.5cm 0cm 2cm 0cm, clip, width=\textwidth]{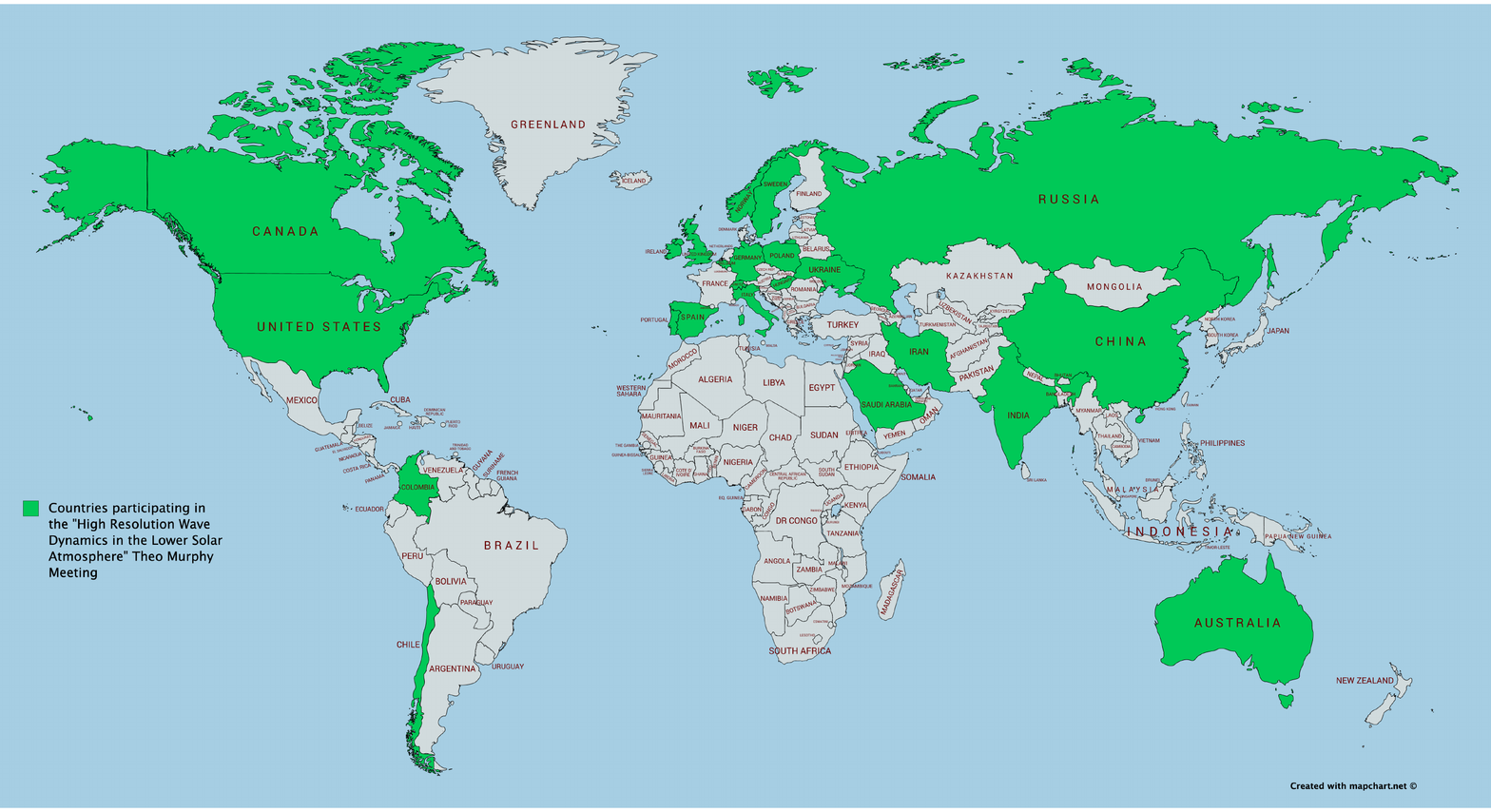}
\caption{A world map, where both the countries of origin and the current institutes of researchers contributing to the workshops and publications featured in the Special Issue of the {\it{Philosophical Transactions of the Royal Society A}} are highlighted in green. The countries highlighted amount to $\sim$45\% of the Earth's total landmass (23 countries in total), showing the widespread influence of research feeding into The Royal Society's Special Issue. Image courtesy of \href{www.mapchart.net}{www.mapchart.net}. }
\label{fig:WaLSAworldmap}
\end{figure}

When assembling both the original WaLSA team and the larger scientific group present at the Theo Murphy Meeting in February~2020, it was paramount to ensure that diversity, in terms of career stage, gender, race, and geographical location, was achieved in order to most readily support the next generation of active solar physics researchers in the field. Of the 28 participants registered for the {\it{``High resolution wave dynamics in the lower solar atmosphere''}} meeting, over 25\% were female and included career stages spanning the very first year of their PhD, through to senior professorial academics at prestigious, globally-recognised universities (see, e.g., those listed in Table~{\ref{tab:WaLSAinstitutes}}). Furthermore, when considering both the countries of origin and the current research institutes of the WaLSA team and Theo Murphy Meeting delegates, 23 countries are incorporated into the worldwide collaborative network provided by the The Royal Society Hooke Committee funding (Figure~{\ref{fig:WaLSAworldmap}}). The widespread nature of WaLSA activities ensures that, at present, approximately 45\% of the Earth's landmass is linked to researchers undertaking cutting-edge studies of waves in the Sun's lower atmosphere through engagement with the WaLSA scheme. However, it can be seen from Figure~{\ref{fig:WaLSAworldmap}} that scientific links with continental Africa are currently missing; something that has been documented across all fields of science in recent years \citep{2019Natur.572..143W}. Thankfully, new initiatives including the UK--South Africa Newton Fund\footnote{\href{https://www.newtonfund.ac.uk/about/about-partner-countries/south-africa/}{https://www.newtonfund.ac.uk/about/about-partner-countries/south-africa/}} and The Royal Society's International Exchange\footnote{\href{https://royalsociety.org/grants-schemes-awards/grants/international-exchanges/}{https://royalsociety.org/grants-schemes-awards/grants/international-exchanges/}} programmes may help facilitate bilateral involvement with developing African research centres, something which the WaLSA team looks forward to actively engaging with over the coming years. 

\section{Publications in the Special Issue}
\label{sec:publicationsinthespecialissue}
The themed Special Issue of the {\it{Philosophical Transactions of the Royal Society A}}, based around the Theo Murphy Meeting {\it{``High resolution wave dynamics in the lower solar atmosphere''}}, incorporates 15 novel publications (excluding this preface) that are at the forefront of current research efforts around the world. Submissions for the Special Issue began in May 2020, with the manuscripts peer reviewed by international experts in the field, alongside scientific and editorial guidance from the Guest Editors Drs. Jess, Keys, Stangalini, and Jafarzadeh. Even with the rigorous peer review process in place, an average submission-to-acceptance time frame of $69 \pm 23$~days was achieved, which recognises the dedication given to the Special Issue by both the authors and international peer reviewers alike. 

In total, the Special Issue comprises of 55 unique authors spanning 34 research institutes, highlighting the global collaborative work undertaken in response to the funding provided by The Royal Society's Hooke Committee. While all of the publications are central to the theme of wave activity in the lower solar atmosphere, some document theoretical and/or numerical findings, while others highlight developments in the observational analyses of high resolution photospheric and/or chromospheric datasets. The subsections below summarise the contributions to the Special Issue by breaking them up into overarching specialist subject areas.

\subsection{Cover Art for the Special Issue}
\label{sec:cover_art}
\begin{figure}[!t]
\centering\includegraphics[width=\textwidth, trim=0cm 12cm 0cm 0cm, clip]{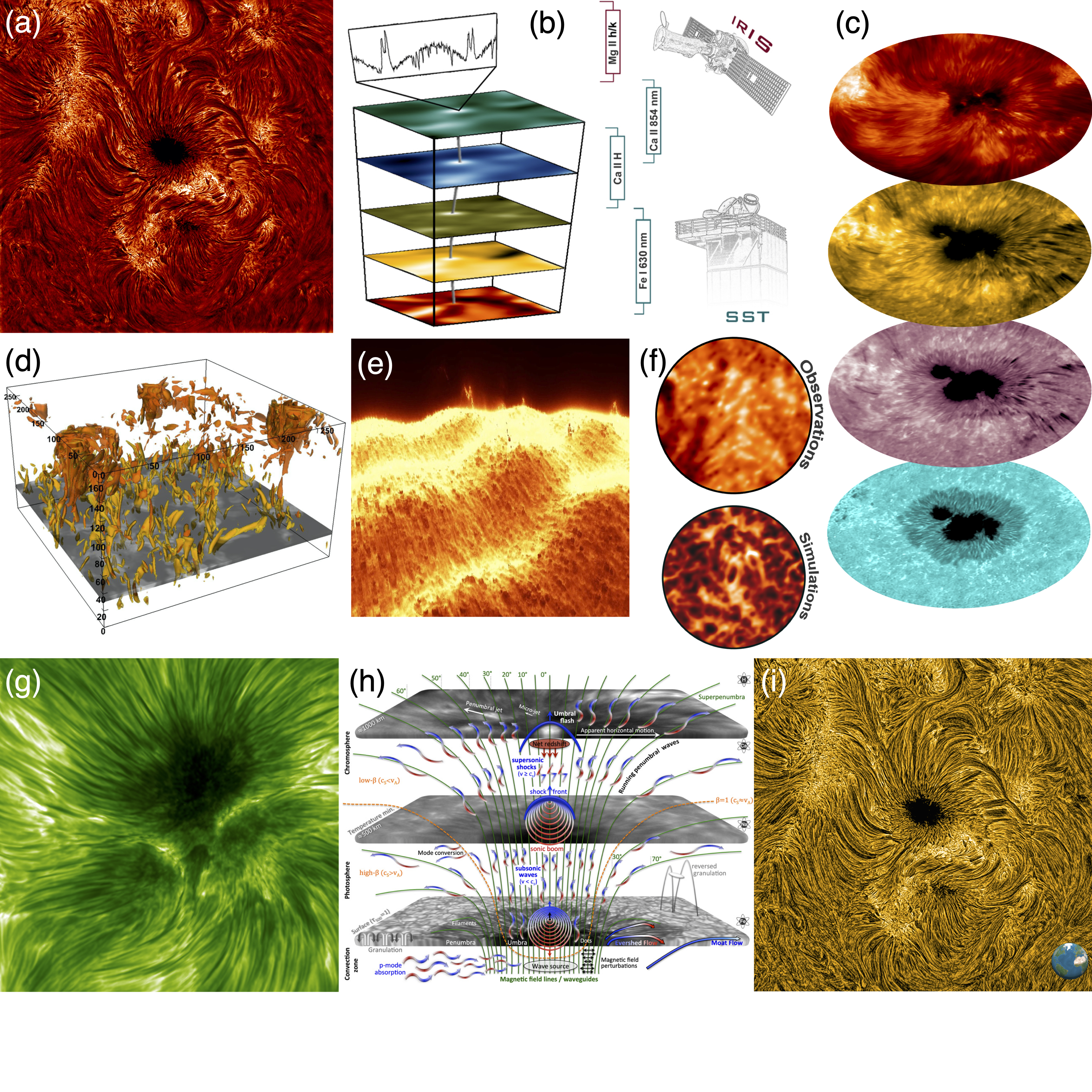}
\caption{All 9 of the potential cover art images that were submitted as part of the public engagement activities described in Section~2{\ref{sec:cover_art}}. Captions for all panels of Figure~{\ref{fig:cover_images}} are displayed in Section~2{\ref{sec:cover_art}}, alongside the corresponding credits for each submission. During the online voting process, over 800 independent votes were registered spanning 18 countries.}
\label{fig:cover_images}
\end{figure}

Due to the large number of publications being written and accepted for the Special Issue of the {\it{Philosophical Transactions of the Royal Society A}}, it was possible to select a single striking, relevant image that encapsulated the {\it{High-resolution wave dynamics in the lower solar atmosphere}} theme of the published manuscripts. Each of the lead authors from the Special Issue were invited to submit (or recommend) an image for consideration that depicted an observational, simulated, and/or artistic impression of wave activity in the lower solar atmosphere. In total, 9 submitted images were received, which are shown in Figure~{\ref{fig:cover_images}}. Captions for each of the 9 potential cover art submissions, alongside the corresponding image credits, are appended below.
\vspace{-2mm}
\begin{enumerate}[label=(\alph*)]
    \item Image of a highly magnetic active region as captured in the chromospheric H$\alpha$ line (656.3~nm) by the Hydrogen-Alpha Rapid Dynamics camera \citep[HARDcam;][]{2012ApJ...757..160J} instrument at the National Science Foundation's Dunn Solar Telescope in New Mexico, USA. The dark sunspot umbra towards the centre of the field-of-view reveals much cooler material as a result of the strong magnetic fields inhibiting the convective mixing of the surrounding plasma. Sunspots play a pivotal role in the guiding of wave motion due to their concentrated magnetic fields. \\ {\it{Image credit: David B. Jess}}.
    \item Multi-layer observations of a small-scale magnetic element through the lower solar atmosphere with the Swedish 1m Solar Telescope (SST) and Interface Region Imaging Spectrograph (IRIS) explorer. For better visibility, the 3D cube has been stretched in height. Swaying motion of the flux tube (due to propagation of MHD transverse waves) and heating signatures (diagnosed as enhanced emission cores in the IRIS Mg~{\sc{ii}}~k spectra) are evident. \\ {\it{Image Credit: Shahin Jafarzadeh and RoCS}}.
    \item A stack of sunspot images captured in the Ca~{\sc{ii}} spectral line at 854.2~nm by the IBIS spectral imaging instrument at the Dunn Solar Telescope in New Mexico, USA. The blue, purple, yellow, and red coloured images sample different parts of the calcium spectral line and reveal the structuring of the sunspot atmosphere from the photosphere (blue) through to the upper chromosphere (red). The dark sunspot umbra towards the centre of the field-of-view reveals much cooler material as a result of the strong magnetic fields inhibiting the convective mixing of the surrounding plasma. \\ {\it{Image credit: David B. Jess}}.
    \item Three-dimensional rendering of the vorticity channels in simulations of solar magneto-convection. Vorticity channels associated with the presence of magnetic flux tubes are highlighted in orange. The grey scale image at the bottom indicates the temperature below the solar surface. \\ {\it{Image credit: Elena Khomenko}}.
    \item An artistic rendering of wave motion in the lower solar atmosphere, where a powerful shader lab in the Unity\footnote{\href{https://unity.com/}{https://unity.com/}} game engine was used to synthesise the solar surface. With DirectX libraries and GPUs it was possible to create the desired wave-like effects to resemble motion in the Sun's atmosphere. \\ {\it{Image credit: Robert Sych and Adam Bielecki \footnote{\href{http://bit.ly/sunshaderunity}{http://bit.ly/sunshaderunity}}}}.
    \item The image illustrates the solar chromosphere at around 1.2~mm wavelength from observations with the Atacama Large Millimeter/submillimeter Array (ALMA; top panel) and from realistic three-dimensional radiation magnetohydrodynamic simulations with the Bifrost code (bottom panel). Various types of waves and oscillations have been observed in these observations and simulations at different spatial scales. \\ {\it{Image Credit: SolarALMA, RoCS and the University of Oslo}}.
    \item Sunspots are large patches of magnetic field that appear prominently on the Sun's surface (photosphere). The image shows the chromosphere of a sunspot, revealing how the magnetic field also shapes the upper layers of the Sun's atmosphere, providing the hair-like structures called fibrils. The image is from the Ca~{\sc{ii}} line (854.2~nm) observed with the Swedish Solar Telescope. \\ {\it{Image Credit: Richard J. Morton and Vasco M. J. Henriques}}.
    \item Wave dynamics in a sunspot observed with the ground-based Dunn Solar Telescope in New Mexico. The simultaneous scanning of multiple spectral lines in the visible to near-infrared range has allowed a fine sampling of the photosphere and chromosphere above the sunspot. Sunspot waves are detected throughout all atmospheric layers, from the lower photosphere to the upper transition region and corona. \\ {\it{Image Credit: Johannes L{\"{o}}hner-B{\"{o}}ttcher\footnote{See the PhD thesis of Johannes L{\"{o}}hner-B{\"{o}}ttcher for further information: {\href{https://freidok.uni-freiburg.de/data/10748}{https://freidok.uni-freiburg.de/data/10748}}}}}.
    \item Same image as panel (a), only now spatially sharpened through use of Multiscale Gaussian Normalisation \cite{2014SoPh..289.2945M}, with a life-size representation of the Earth depicted in the lower-right to provide a sense of scale. \\ {\it{Image credit: David B. Jess}}.
\end{enumerate}

In order to promote engagement with the general public, the 9 submitted images were uploaded to a voting platform accessible by all academic staff, researchers, scientists, and members of the public. Links to the voting webpage were circulated via social media (Facebook, Instagram, Twitter, etc.) posts in order to reach maximum visibility around the globe. The voting process was open across a period of 2 weeks, and the image with the highest overall rating was selected for the cover art of the Special Issue of the {\it{Philosophical Transactions of the Royal Society A}}. 

\begin{figure}[!t]
\centering\includegraphics[width=\textwidth]{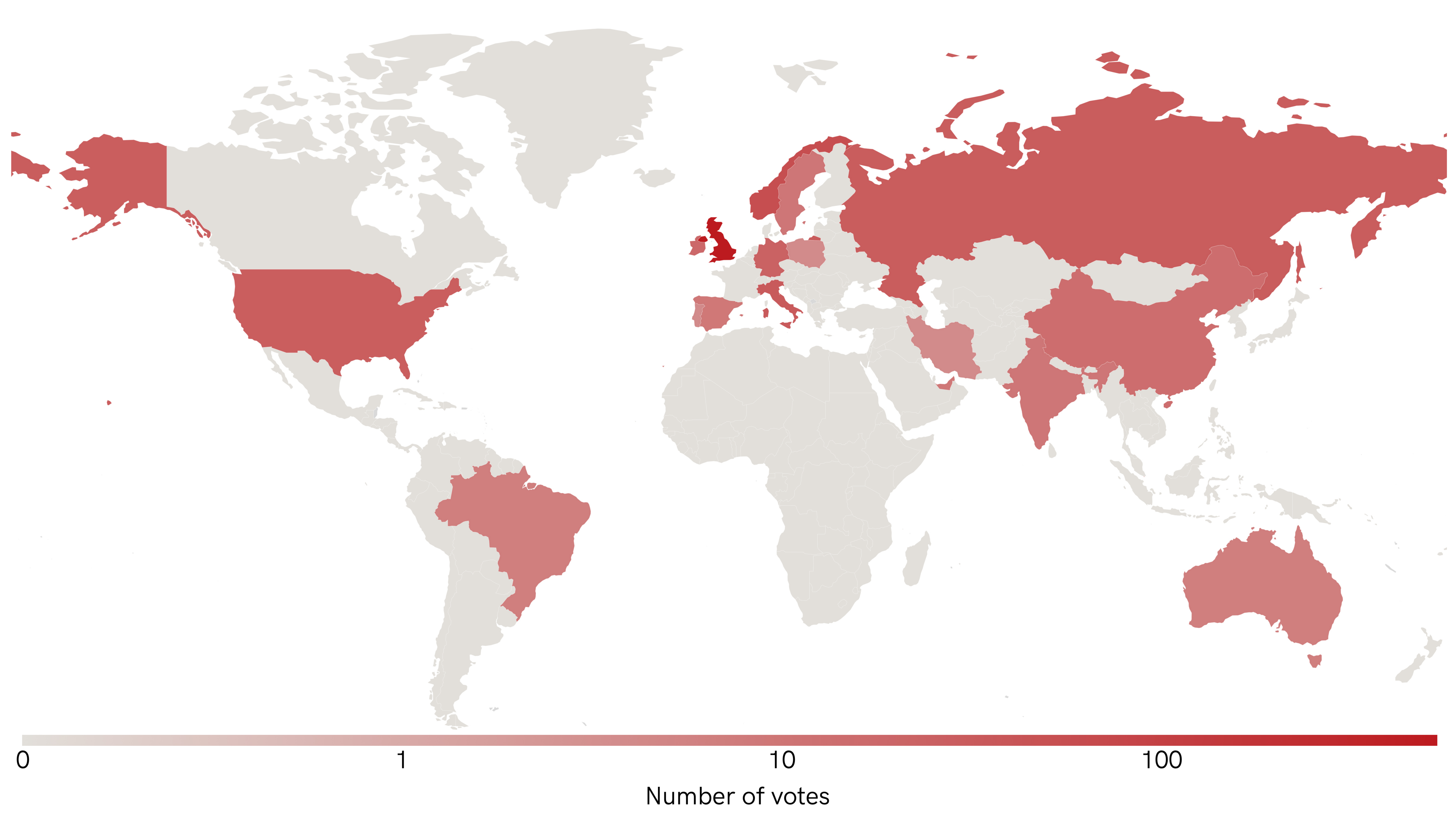}
\caption{World map depicting the number of votes registered per country as part of the cover art selection process, with the white-to-red colour bar placed on a log-scale to better reveal countries with small voting statistics. Image data compiled by Shahin Jafarzadeh, displayed courtesy of \href{https://www.amcharts.com}{https://www.amcharts.com}, and based on the public voting platform produced by the WaLSA team.}
\label{fig:heatmap}
\end{figure}

Following the 2 weeks of the public vote on potential cover art images, over 800 independent votes were logged. While the specific location of the voters remained anonymous, we were able to log the countries of origin for each vote. In total, 18 countries submitted votes, incorporating (in alphabetical order) Australia, Brazil, China, Germany, India, Iran, Ireland, Italy, Norway, Poland, Portugal, Russia, Singapore, Spain, Sweden, United Arab Emirates, United Kingdom, and the United States. The most and least active countries during the voting process were the United Kingdom (541 votes) and Poland (4 votes), respectively. To better visualise the registered votes, a `heat map' was created to show the widespread engagement that was possible with such online activities (Figure~{\ref{fig:heatmap}}), including participation from members of the general public. It can be seen from Figure~{\ref{fig:heatmap}} that countries not directly affiliated with the Theo Murphy meeting (e.g., those displayed in Figure~{\ref{fig:WaLSAworldmap}}) were still able to participate in the voting process; highlighting the power of modern social media platforms. 

\subsection{Simulations of the Solar Atmosphere}
Starting with high resolution simulations of the base of the photosphere, Fleck et~al. \citep{Bernhard} examines and compares wave propagation characteristics across a number of internationally recognised magnetohydrodynamic (MHD) codes. The authors find troubling conclusions, in that not all MHD simulations display consistent wave signatures, highlighting the need for future collaborative efforts to ensure wave phenomena at the base of the visible photosphere is accurately and consistently replicated in these prevalent simulations. Introducing more concentrated magnetic fields to the simulation domain in the form of a magnetic flux sheet, Keys et~al. \citep{Peter} investigates the accuracy of Stokes inversions for prominent photospheric line profiles, including the 6301{\,}{\AA} and 6302{\,}{\AA} line pair, and establishes whether accurate atmospheric parameters can be returned for typical spatial resolutions synonymous with modern (and upcoming) ground-based observatories. The authors find that the inversions they employ return temperatures, line-of-sight magnetic fields, and line-of-sight velocities comparable to the original input numerical atmosphere, even with spectral asymmetries introduced from the generation and propagation of waves in the atmosphere. These are exciting results, since it means the improved spatial, temporal, and spectral resolutions expected from the next generation of ground-based and space-borne telescopes will be able to accurately uncover and characterise wave phenomena on the smallest spatial scales achieved to date. 

As waves continue to propagate into the upper solar photosphere and chromosphere, the associated plasma motions encounter the increasing influences of ambipolar and Hall effects. Khomenko et~al. \citep{Elena} examine how the ambipolar and Hall terms affect the vorticities seen in three-dimensional simulations of magneto-convection. The authors find that ambipolar diffusion results in a reduction of vorticity in the upper chromosphere, with vortical perturbations being dissipated and converted into thermal energy. Conversely, the authors find that the Hall effect acts to strongly enhance the vorticity. When the Hall effect is included, the authors find that the magnetic field is more vertical in nature and forms more flux-tube like structures that are long-lived, with the enhanced Hall effect possibly stabilising the tubes against convective motions. These vertical flux tubes act as funnels channelling vortical motions to upper layers, which are dissipated leading to temperature increases in the chromosphere that can endure for a significant amount of time.

 Once waves, in particular slow magnetoacoustic modes, begin to traverse the solar chromosphere, they often steepen into local shock fronts, manifesting as intensely heated concentrations of plasma that outline the motion path of the wave. Through examination of cutting-edge radiative MHD Bifrost simulations, Eklund et~al. \citep{Henrik} investigate the evolution of brightness temperatures coupled to a strong developing shock wave. The authors demonstrate that millimetre brightness temperatures \citep[i.e., consistent with radio observations from the Atacama Large Millimeter/sub-millimeter Array (ALMA);][]{2009IEEEP..97.1463W, 2016SSRv..200....1W} efficiently track upwardly propagating shock waves in the middle chromosphere, which provides an important diagnostic tool to better understand small-scale dynamics across these turbulent regions of the Sun's atmosphere.

\subsection{Gravity Waves}
In relatively weak magnetic regions of the lower solar atmosphere, gravity waves with frequencies shorter than $\lesssim 2$~mHz (periods $\gtrsim 500$~s) have been identified in both observations and simulations \citep{2010MNRAS.402..386N,2011A&A...532A.111K,2017ApJ...835..148V,2020A&A...633A.140V}. Such waves have been found to have larger energy flux, by approximately one order of magnitude, compared to co-spatial acoustic waves \citep{2008ApJ...681L.125S}, thus can potentially contribute towards heating of the chromosphere and beyond. Detection of the gravity waves in the solar atmosphere has, however, been a challenging task. Using a three dimensional cross-correlation technique involving time series of vertical velocities measured at different atmospheric heights, Calchetti et~al. \citep{Daniele} propose a new approach to directly identify and characterise gravity waves in the lower solar atmosphere. The authors further validate their method using an observational dataset as well as a simple numerical model.

Internal gravity waves are mainly driven by convective updrafts reaching the stratified atmosphere above, with low frequency ultraviolet brightness fluctuations believed to be the result of these waves dissipating in higher regions \citep{Rutten2003}. The role of magnetic fields in the propagation of internal gravity waves is less clear. Likewise, the link between brightness fluctuations in internetwork regions and internal gravity waves is not fully resolved. Using three-dimensional simulations of the near-surface obtained with the CO$^{5}$BOLD code \citep{cobold}, Vigeesh et~al. \citep{Gangadharan} find that in low layers the magnetic field does not play a significant role in the propagation properties of internal gravity waves. However, the authors find that in higher layers (above $\sim$0.4~Mm) the average magnetic field orientation dictates the propagation properties, with internal gravity waves reflected downwards in predominately vertical field models while propagating freely upwards in horizontal and non-magnetic models. The authors also suggest that wave-breaking in internetwork models may result in the ultraviolet brightness fluctuations reported in the literature.

\subsection{Pioneering ALMA Radio Observations}
Transitioning into radio observations of the solar atmosphere, Jafarzadeh et~al. \citep{Shahin} and Guevara~G{\'{o}}mez et~al. \citep{Juan} study pioneering ALMA observations to search for wave signatures in brightness temperature values that are coupled to the chromospheric plasma. Examining the brightest (and therefore hottest) locations of a plage/enhanced-network region observed by ALMA, Guevara~G{\'{o}}mez et~al. \citep{Juan} uncover significant oscillations in brightness temperature, size, and horizontal velocity, which are believed to be combined signatures of fast sausage-mode and Alfv{\'{e}}nic waves in these small structures. Turning attention towards ALMA observations demonstrating varying degrees of magnetic complexity, Jafarzadeh et~al. \citep{Shahin} highlights the potential challenges when interpreting radio observations. Notably, the authors find that the wave power spectra vary from dataset to dataset, which is further examined in terms of the underlying magnetic topology. Importantly, Jafarzadeh et~al. \citep{Shahin} propose that more theoretical and modelling work is required to understand why only a very small fraction of the ALMA data examined show peak power near 5.5~mHz, which is in contradiction to other chromospheric diagnostics where very pronounced oscillations are regularly found in this band. 

\subsection{Observations of Large-scale Solar Magnetism}
As the concentrations of solar magnetism continue to grow, larger-scale structures begin to form in the Sun's atmosphere, including the formation of solar pores. These highly magnetic elements are studied by Gilchrist-Millar et~al. \citep{Caitlin}, who captured photospheric spectropolarimetric profiles of the infrared Si~{\sc{i}}~10{\,}827~{\AA} absorption line for a collection of pores located with the same field-of-view of the Dunn Solar Telescope. The authors employ polarimetric inversion routines and high time resolution data products to examine the height dependency of wave energy flux in the lower solar atmosphere. Importantly, Gilchrist-Millar et~al. \citep{Caitlin} discover that while pore structures act as efficient wave conduits to guide energy flux into the outer regions of the Sun's atmosphere, the photospheric layers rapidly damp the embedded wave energy by several orders of magnitude. These results, including the short damping lengths uncovered, pose important questions for next-generation facilities that should be able to probe such features with unprecedented polarimetric precision.

Fully formed sunspots are a visually striking manifestation of concentrated magnetic fields in the Sun's atmosphere. Through novel filtering techniques, Sych et~al. \citep{Robert} examine how wave activity in the most intensely magnetic umbral locations allows the embedded fine structure of the sunspot to be probed with high levels of precision. The authors find that the motion path of visible wave trains occurs along preferential directions depending upon the magnetic field inclination angles present within the umbra. Sych et~al. \citep{Robert} propose that small-scale umbral structuring may be responsible for the creation of ubiquitous umbral flashes, and highlight the important need for next-generation larger-aperture solar telescopes to unequivocally answer this question. In keeping with umbral wave observations, Albidah et~al. \citep{Abdulrahman} examine a sequence of high-cadence H$\alpha$ images that captured a near circularly symmetric sunspot using the HARDcam instrument on the Dunn Solar Telescope. Applying cutting-edge Proper Orthogonal Decomposition (POD) and Dynamic Mode Decomposition (DMD) techniques, the authors are able to identify low-order MHD wave modes as coherent oscillations across the entire sunspot umbra. Through examination of spatial and temporal orthogonality, Albidah et~al. \citep{Abdulrahman} provide evidence to support the co-existence of sausage {\it{and}} kink modes within the same magnetic structure, which demonstrates the suitability of their techniques to more widespread solar observations of sunspot waves.

One of the largest and most magnetically intense sunspots of the previous solar cycle passed through the Earth's line-of-sight during May 2016, and was captured by the Interferometric BIdimensional Spectrometer \citep[IBIS;][]{2006SoPh..236..415C} Fabry-P{\'{e}}rot instrument at the Dunn Solar Telescope. Utilising chromospheric Ca~{\sc{ii}}~8542~{\AA} spectropolarimetric imaging scans, Stangalini et~al. \citep{Marco} investigates the relationship between intensity and circular polarisation fluctuations observed within the large-scale sunspot. The authors find clear correlations between the intensities and magnetic signals, suggesting the detection of Alfv{\'{e}}nic perturbations. This has important implications for investigations of chromospheric mode conversion physics \citep{2018NatPh..14..480G, 2020ApJ...892...49H}, in addition to shining light on the role of magnetic waves in the generation of the First Ionisation Potential (FIP) effect that is ubiquitously observed in the solar corona, in the solar wind, and solar-like stellar coronae \citep{2015LRSP...12....2L}, and was therefore proposed as a potential mechanism to trace back the solar wind to its atmospheric source. 

The dynamics displayed in high-resolution spectral sunspot observations often creates challenging problems for researchers attempting to understand the mechanisms responsible for generating asymmetric, centrally reversed, and/or heavily skewed line profiles. The interplay between source functions and opacity effects often requires the fitment of two (or more) profiles to any given spectral line in order to best constrain the underlying dynamics. This is a time consuming endeavour to perform manually, prompting MacBride et~al. \citep{Conor} to apply machine learning techniques to this challenging problem to accurately, efficiently, and repeatedly extract the key components of the observed line profiles. The authors employed a training dataset based on IBIS Ca~{\sc{ii}}~8542~{\AA} sunspot observations, and when applied to high-cadence spectral scans it was able to accurately fit over 600{\,}000 spectral profiles in approximately 2~hours, with the potential to further parallelise via multiple CPU cores and/or the implementation of GPUs. MacBride et~al. \citep{Conor} highlight the potential of their techniques to be applied to upcoming observations from the National Science Foundation's Daniel K. Inouye Solar Telescope \citep[DKIST;][]{2016AN....337.1064T} in real-time, thus providing important wave diagnostics live at the observing site to help pinpoint regions of interest for specific wave studies. 

As observations move away from the umbral core of a sunspot, the magnetic field lines begin to more heavily incline with respect to the solar normal. Using high-resolution Ca~{\sc{ii}}~8542~{\AA} observations from the Swedish 1m Solar Telescope, Morton et~al. \citep{Richard} examine chromospheric super-penumbral fibrils that span out radially from the underlying sunspot. Through extensive statistical analyses, the authors find that these narrow structures show ubiquitous wave-like motions that can be interpreted as MHD kink waves, which are broadly consistent with previous observations of chromospheric transverse waves in quiet Sun fibrils. Morton et~al. \citep{Richard} speculate that these motions may be able to improve our understanding of how MHD wave energy is transferred through the atmosphere of a sunspot. Zooming out to incorporate the full extent of a sunspot active region, the larger-scale interactions of the embedded magnetic fields provides a number of distinct locations from which unique wave relationships can be examined. One such example involves the formation of oscillatory phenomena in the vicinity of a polarity inversion line. Norton et~al. \citep{Aimee} harness the large field of view of the Helioseismic Magnetic Imager \citep[HMI;][]{2012SoPh..275..207S} onboard SDO to examine the amplitudes and phase relations of oscillations existing in quiet Sun, plage, umbral, and polarity inversion line (PIL) locations of an active region. The authors derived values are consistent with slow standing or fast standing surface sausage mode waves, leading Norton et~al. \citep{Aimee} to propose that line width variations (alongside their phase relations with intensity and magnetic oscillations) may offer a novel new way to further differentiate wave mode mechanics in the solar atmosphere. Importantly, Norton et~al. \citep{Aimee} reveal evidence for the mode conversion of Doppler oscillations into magnetic waves, highlighting the important interplay of MHD waves across all layers of the Sun's atmosphere. 

\section{Conclusions and Future Directions}
\label{sec:conclusionsandfuturedirections}
Solar physics research is ideally positioned to make rapid discoveries over the coming decades, with the imminent full-scale operation of the 4m DKIST facility, in addition to ongoing work linked to the development of the 2m Indian National Large Solar Telescope \citep[NLST;][]{2010AN....331..628H} and the 4m European Solar Telescope \citep[EST;][]{2010AN....331..615C}. In addition to these ground-based observatories, next-generation balloon-based and space-borne facilities, including the continually updated {\sc Sunrise} observatory \citep[][]{2010ApJ...723L.127S, 2017ApJS..229....2S}, Solar Orbiter \citep{2020A&A...642A...1M}, and the JAXA M-class Solar-C \citep{2019SPIE11118E..07S} future mission, will help probe the photospheric and chromospheric interactions with other regions of the solar atmosphere on unprecedented scales. In particular, novel instrumentation, including fibre-fed imaging spectrographs \citep[e.g., the Diffraction-Limited Near-Infrared Spectropolarimeter, DL-NIRSP;][]{2019AAS...23410612J}, will provide the community with the much anticipated spectral, polarimetric, spatial, and temporal resolutions necessary to accurately constrain the rapidly evolving, yet often small amplitude fluctuations associated with wave phenomena in the Sun's atmosphere. 

In addition to the observational improvements coming online over the next decade, researchers will also be able to capitalise on next-generation high performance computing (HPC) infrastructures that are currently being built and updated. Within the UK, the Distributed Research utilising Advanced Computing\footnote{\href{https://dirac.ac.uk/}{https://dirac.ac.uk/}} (DiRAC) and ARCHER\footnote{\href{http://www.archer.ac.uk/}{http://www.archer.ac.uk/}} HPC systems provide over 6~petaflops of compute capability, while international HPC services such as La Red Espa{\~{n}}ola de Supercomputaci{\'{o}}n\footnote{\href{https://www.res.es/en}{https://www.res.es/en}} (RES, Spanish Supercomputing Network), the Norwegian academic e-infrastructure\footnote{\href{https://www.sigma2.no/systems}{https://www.sigma2.no/systems}}, and the NASA Pleiades\footnote{\href{https://www.nas.nasa.gov/hecc/resources/pleiades.html}{https://www.nas.nasa.gov/hecc/resources/pleiades.html}} supercomputer can boost the compute capabilities up into the tens of petaflops. Such evolving HPC facilities are crucial for the accurate replication of physics, particularly down to the spatial and temporal scales imminently visible by the newest observing facilities. 

The publications presented in the current Special Issue demonstrate the computational and observational challenges that await the widespread community. However, it is clear that the next generation of committed solar physicists, alongside the friendly, inclusive, collaborative environment and knowledge base brought by senior researchers and academics, can readily foster solutions to any potential problems raised in the field of oscillations and waves in the lower solar atmosphere. We therefore look forward in anticipation to the new era of understanding that will be brought to fruition by the newest researchers, observatories, and computing facilities over the decades to come. 

\vspace{10mm}
\aucontribute{DBJ, PHK, MS and SJ acted as Guest Editors for the {\it{``High-resolution wave dynamics in the lower solar atmosphere''}} Special Issue of the {\it{Philosophical Transactions of the Royal Society A}}. All authors drafted, read and approved this introduction.}

\vspace{5mm}
\competing{The authors declare that they have no competing interests.}

\vspace{5mm}
\funding{Funding has come from the following sources: 
\vspace{-3mm}
\begin{itemize}
	\item UK Science and Technology Facilities Council (STFC) grants (ST/K004220/1, ST/L002744/1 \& ST/T00021X/1);
    \item Invest NI and Randox Laboratories Ltd. Research \& Development Grant (059RDEN-1);
    \item European Union's Horizon 2020 research and innovation programme (grant agreement no. 682462);
    \item Research Council of Norway through its Centres of Excellence scheme (project no. 262622);
    \item European Union's Horizon 2020 research and innovation programme (grant agreement No 724326);
    \item Research Council of Norway (project number 262622); and
    \item The Royal Society (grant Hooke18b/SCTM).
\end{itemize}}

\ack{
DBJ would like to thank the UK Science and Technology Facilities Council (STFC) for an Ernest Rutherford Fellowship (ST/K004220/1), in addition to dedicated standard and consolidated grants (ST/L002744/1 and ST/T00021X/1) that allowed this project to be started.
DBJ also wishes to thank Invest NI and Randox Laboratories Ltd. for the award of a Research and Development Grant (059RDEN-1) that supported the development of computational techniques. 
SJ acknowledges support from the European Research Council under the European Union's Horizon 2020 research and innovation programme (grant agreement no. 682462) and from the Research Council of Norway through its Centres of Excellence scheme (project no. 262622).
We wish to acknowledge scientific discussions with the Waves in the Lower Solar Atmosphere (WaLSA; \href{https://www.WaLSA.team}{www.WaLSA.team}) team, which is supported by the Research Council of Norway (project no. 262622) and the Royal Society (award no. Hooke18b/SCTM).
We are also grateful to the Royal Society staff, including Alice~Power, Chloe~Mavrommatis, Yan~Zhao, Annabel~Sturgess, and Amie~Mustill for their guidance, advice, and patience when helping us plan the Theo Murphy meeting and editing the current Special Issue. 
}


\bibliographystyle{rstasj}

\begin{thebibliography}{57}
\providecommand{\natexlab}[1]{#1}
\providecommand{\url}[1]{\texttt{#1}}
\expandafter\ifx\csname urlstyle\endcsname\relax
  \providecommand{\doi}[1]{doi: #1}\else
  \providecommand{\doi}{doi: \begingroup \urlstyle{rm}\Url}\fi

\bibitem[{Leighton} et~al.(1962){Leighton}, {Noyes}, and
  {Simon}]{1962ApJ...135..474L}
{Leighton} RB, {Noyes} RW, {Simon} GW.
\newblock 1962, {Velocity Fields in the Solar Atmosphere. I. Preliminary
  Report.}
\newblock \emph{\apj}, 135:\penalty0 474.
\newblock (\doi{10.1086/147285})

\bibitem[{Noyes} and {Leighton}(1963)]{1963ApJ...138..631N}
{Noyes} RW, {Leighton} RB.
\newblock 1963, {Velocity Fields in the Solar Atmosphere. II. The Oscillatory
  Field.}
\newblock \emph{\apj}, 138:\penalty0 631.
\newblock (\doi{10.1086/147675})

\bibitem[{Jess} et~al.(2015){Jess}, {Morton}, {Verth}, {Fedun}, {Grant}, and
  {Giagkiozis}]{2015SSRv..190..103J}
{Jess} DB, {Morton} RJ, {Verth} G, {Fedun} V, {Grant} SDT, {Giagkiozis} I.
\newblock 2015, {Multiwavelength Studies of MHD Waves in the Solar
  Chromosphere. An Overview of Recent Results}.
\newblock \emph{\ssr}, 190\penalty0 (1-4):\penalty0 103--161.
\newblock (\doi{10.1007/s11214-015-0141-3})

\bibitem[{Dunn}(1969)]{1969S&T....38..368D}
{Dunn} RB.
\newblock 1969, {Sacramento Peak's New Solar Telescope}.
\newblock \emph{\skytel}, 38:\penalty0 368

\bibitem[{Scharmer} et~al.(2003){Scharmer}, {Bjelksjo}, {Korhonen}, {Lindberg},
  and {Petterson}]{2003SPIE.4853..341S}
{Scharmer} GB, {Bjelksjo} K, {Korhonen} TK, {Lindberg} B, {Petterson} B.
\newblock 2003,
\newblock , \emph{\procspie}, 341-350volume 4853 of \emph{Society of
  Photo--Optical Instrumentation Engineers (SPIE) Conference Series}pages.
\newblock (\doi{10.1117/12.460377})

\bibitem[{Cao} et~al.(2010){Cao}, {Gorceix}, {Coulter}, {Ahn}, {Rimmele}, and
  {Goode}]{2010AN....331..636C}
{Cao} W, {Gorceix} N, {Coulter} R, {Ahn} K, {Rimmele} TR, {Goode} PR.
\newblock 2010, {Scientific instrumentation for the 1.6 m New Solar Telescope
  in Big Bear}.
\newblock \emph{Astronomische Nachrichten}, 331\penalty0 (6):\penalty0 636.
\newblock (\doi{10.1002/asna.201011390})

\bibitem[{Kosugi} et~al.(2007){Kosugi}, {Matsuzaki}, {Sakao}, {Shimizu},
  {Sone}, {Tachikawa}, {Hashimoto}, {Minesugi}, {Ohnishi}, {Yamada}, {Tsuneta},
  {Hara}, {Ichimoto}, {Suematsu}, {Shimojo}, {Watanabe}, {Shimada}, {Davis},
  {Hill}, {Owens}, {Title}, {Culhane}, {Harra}, {Doschek}, and
  {Golub}]{2007SoPh..243....3K}
{Kosugi} T, {Matsuzaki} K, {Sakao} T, {Shimizu} T, {Sone} Y, {Tachikawa} S,
  {Hashimoto} T, {Minesugi} K, {Ohnishi} A, {Yamada} T, {Tsuneta} S, {Hara} H,
  {Ichimoto} K, {Suematsu} Y, {Shimojo} M, {Watanabe} T, {Shimada} S, {Davis}
  JM, {Hill} LD, {Owens} JK, {Title} AM, {Culhane} JL, {Harra} LK, {Doschek}
  GA, {Golub} L.
\newblock 2007, {The Hinode (Solar-B) Mission: An Overview}.
\newblock \emph{\solphys}, 243\penalty0 (1):\penalty0 3--17.
\newblock (\doi{10.1007/s11207-007-9014-6})

\bibitem[{De Pontieu} et~al.(2014{\natexlab{a}}){De Pontieu}, {Title}, {Lemen},
  {Kushner}, {Akin}, {Allard}, {Berger}, {Boerner}, {Cheung}, {Chou}, {Drake},
  {Duncan}, {Freeland}, {Heyman}, {Hoffman}, {Hurlburt}, {Lindgren}, {Mathur},
  {Rehse}, {Sabolish}, {Seguin}, {Schrijver}, {Tarbell}, {W{\"u}lser},
  {Wolfson}, {Yanari}, {Mudge}, {Nguyen-Phuc}, {Timmons}, {van Bezooijen},
  {Weingrod}, {Brookner}, {Butcher}, {Dougherty}, {Eder}, {Knagenhjelm},
  {Larsen}, {Mansir}, {Phan}, {Boyle}, {Cheimets}, {DeLuca}, {Golub}, {Gates},
  {Hertz}, {McKillop}, {Park}, {Perry}, {Podgorski}, {Reeves}, {Saar}, {Testa},
  {Tian}, {Weber}, {Dunn}, {Eccles}, {Jaeggli}, {Kankelborg}, {Mashburn},
  {Pust}, {Springer}, {Carvalho}, {Kleint}, {Marmie}, {Mazmanian}, {Pereira},
  {Sawyer}, {Strong}, {Worden}, {Carlsson}, {Hansteen}, {Leenaarts},
  {Wiesmann}, {Aloise}, {Chu}, {Bush}, {Scherrer}, {Brekke}, {Martinez-Sykora},
  {Lites}, {McIntosh}, {Uitenbroek}, {Okamoto}, {Gummin}, {Auker}, {Jerram},
  {Pool}, and {Waltham}]{2014SoPh..289.2733D}
{De Pontieu} B, {Title} AM, {Lemen} JR, {Kushner} GD, {Akin} DJ, {Allard} B,
  {Berger} T, {Boerner} P, {Cheung} M, {Chou} C, {Drake} JF, {Duncan} DW,
  {Freeland} S, {Heyman} GF, {Hoffman} C, {Hurlburt} NE, {Lindgren} RW,
  {Mathur} D, {Rehse} R, {Sabolish} D, {Seguin} R, {Schrijver} CJ, {Tarbell}
  TD, {W{\"u}lser} JP, {Wolfson} CJ, {Yanari} C, {Mudge} J, {Nguyen-Phuc} N,
  {Timmons} R, {van Bezooijen} R, {Weingrod} I, {Brookner} R, {Butcher} G,
  {Dougherty} B, {Eder} J, {Knagenhjelm} V, {Larsen} S, {Mansir} D, {Phan} L,
  {Boyle} P, {Cheimets} PN, {DeLuca} EE, {Golub} L, {Gates} R, {Hertz} E,
  {McKillop} S, {Park} S, {Perry} T, {Podgorski} WA, {Reeves} K, {Saar} S,
  {Testa} P, {Tian} H, {Weber} M, {Dunn} C, {Eccles} S, {Jaeggli} SA,
  {Kankelborg} CC, {Mashburn} K, {Pust} N, {Springer} L, {Carvalho} R, {Kleint}
  L, {Marmie} J, {Mazmanian} E, {Pereira} TMD, {Sawyer} S, {Strong} J, {Worden}
  SP, {Carlsson} M, {Hansteen} VH, {Leenaarts} J, {Wiesmann} M, {Aloise} J,
  {Chu} KC, {Bush} RI, {Scherrer} PH, {Brekke} P, {Martinez-Sykora} J, {Lites}
  BW, {McIntosh} SW, {Uitenbroek} H, {Okamoto} TJ, {Gummin} MA, {Auker} G,
  {Jerram} P, {Pool} P, {Waltham} N.
\newblock 2014{\natexlab{a}}, {The Interface Region Imaging Spectrograph
  (IRIS)}.
\newblock \emph{\solphys}, 289\penalty0 (7):\penalty0 2733--2779.
\newblock (\doi{10.1007/s11207-014-0485-y})

\bibitem[{Tritschler} et~al.(2016){Tritschler}, {Rimmele}, {Berukoff},
  {Casini}, {Kuhn}, {Lin}, {Rast}, {McMullin}, {Schmidt}, {W{\"o}ger}, and
  {DKIST Team}]{2016AN....337.1064T}
{Tritschler} A, {Rimmele} TR, {Berukoff} S, {Casini} R, {Kuhn} JR, {Lin} H,
  {Rast} MP, {McMullin} JP, {Schmidt} W, {W{\"o}ger} F, {DKIST Team}.
\newblock 2016, {Daniel K. Inouye Solar Telescope: High-resolution observing of
  the dynamic Sun}.
\newblock \emph{Astronomische Nachrichten}, 337\penalty0 (10):\penalty0 1064.
\newblock (\doi{10.1002/asna.201612434})

\bibitem[{De Pontieu} et~al.(2014{\natexlab{b}}){De Pontieu}, {Rouppe van der
  Voort}, {McIntosh}, {Pereira}, {Carlsson}, {Hansteen}, {Skogsrud}, {Lemen},
  {Title}, {Boerner}, {Hurlburt}, {Tarbell}, {Wuelser}, {De Luca}, {Golub},
  {McKillop}, {Reeves}, {Saar}, {Testa}, {Tian}, {Kankelborg}, {Jaeggli},
  {Kleint}, and {Martinez-Sykora}]{2014Sci...346D.315D}
{De Pontieu} B, {Rouppe van der Voort} L, {McIntosh} SW, {Pereira} TMD,
  {Carlsson} M, {Hansteen} V, {Skogsrud} H, {Lemen} J, {Title} A, {Boerner} P,
  {Hurlburt} N, {Tarbell} TD, {Wuelser} JP, {De Luca} EE, {Golub} L, {McKillop}
  S, {Reeves} K, {Saar} S, {Testa} P, {Tian} H, {Kankelborg} C, {Jaeggli} S,
  {Kleint} L, {Martinez-Sykora} J.
\newblock 2014{\natexlab{b}}, {On the prevalence of small-scale twist in the
  solar chromosphere and transition region}.
\newblock \emph{Science}, 346\penalty0 (6207):\penalty0 1255732.
\newblock (\doi{10.1126/science.1255732})

\bibitem[{Mart{\'\i}nez-Sykora} et~al.(2017){Mart{\'\i}nez-Sykora}, {De
  Pontieu}, {Hansteen}, {Rouppe van der Voort}, {Carlsson}, and
  {Pereira}]{2017Sci...356.1269M}
{Mart{\'\i}nez-Sykora} J, {De Pontieu} B, {Hansteen} VH, {Rouppe van der Voort}
  L, {Carlsson} M, {Pereira} TMD.
\newblock 2017, {On the generation of solar spicules and Alfv{\'e}nic waves}.
\newblock \emph{Science}, 356\penalty0 (6344):\penalty0 1269--1272.
\newblock (\doi{10.1126/science.aah5412})

\bibitem[{Srivastava} et~al.(2017){Srivastava}, {Shetye}, {Murawski}, {Doyle},
  {Stangalini}, {Scullion}, {Ray}, {W{\'o}jcik}, and
  {Dwivedi}]{2017NatSR...743147S}
{Srivastava} AK, {Shetye} J, {Murawski} K, {Doyle} JG, {Stangalini} M,
  {Scullion} E, {Ray} T, {W{\'o}jcik} DP, {Dwivedi} BN.
\newblock 2017, {High-frequency torsional Alfv{\'e}n waves as an energy source
  for coronal heating}.
\newblock \emph{Nature Scientific Reports}, 7:\penalty0 43147.
\newblock (\doi{10.1038/srep43147})

\bibitem[{Grant} et~al.(2018){Grant}, {Jess}, {Zaqarashvili}, {Beck},
  {Socas-Navarro}, {Aschwanden}, {Keys}, {Christian}, {Houston}, and
  {Hewitt}]{2018NatPh..14..480G}
{Grant} SDT, {Jess} DB, {Zaqarashvili} TV, {Beck} C, {Socas-Navarro} H,
  {Aschwanden} MJ, {Keys} PH, {Christian} DJ, {Houston} SJ, {Hewitt} RL.
\newblock 2018, {Alfv{\'e}n wave dissipation in the solar chromosphere}.
\newblock \emph{Nature Physics}, 14\penalty0 (5):\penalty0 480--483.
\newblock (\doi{10.1038/s41567-018-0058-3})

\bibitem[{Samanta} et~al.(2019){Samanta}, {Tian}, {Yurchyshyn}, {Peter}, {Cao},
  {Sterling}, {Erd{\'e}lyi}, {Ahn}, {Feng}, {Utz}, {Banerjee}, and
  {Chen}]{2019Sci...366..890S}
{Samanta} T, {Tian} H, {Yurchyshyn} V, {Peter} H, {Cao} W, {Sterling} A,
  {Erd{\'e}lyi} R, {Ahn} K, {Feng} S, {Utz} D, {Banerjee} D, {Chen} Y.
\newblock 2019, {Generation of solar spicules and subsequent atmospheric
  heating}.
\newblock \emph{Science}, 366\penalty0 (6467):\penalty0 890--894.
\newblock (\doi{10.1126/science.aaw2796})

\bibitem[{Jess} et~al.(2020{\natexlab{a}}){Jess}, {Snow}, {Houston}, {Botha},
  {Fleck}, {Krishna Prasad}, {Asensio Ramos}, {Morton}, {Keys}, {Jafarzadeh},
  {Stangalini}, {Grant}, and {Christian}]{2020NatAs...4..220J}
{Jess} DB, {Snow} B, {Houston} SJ, {Botha} GJJ, {Fleck} B, {Krishna Prasad} S,
  {Asensio Ramos} A, {Morton} RJ, {Keys} PH, {Jafarzadeh} S, {Stangalini} M,
  {Grant} SDT, {Christian} DJ.
\newblock 2020{\natexlab{a}}, {A chromospheric resonance cavity in a sunspot
  mapped with seismology}.
\newblock \emph{Nature Astronomy}, 4:\penalty0 220--227.
\newblock (\doi{10.1038/s41550-019-0945-2})

\bibitem[{Jess} et~al.(2020{\natexlab{b}}){Jess}, {Snow}, {Fleck},
  {Stangalini}, and {Jafarzadeh}]{2020NatAs.tmp..149J}
{Jess} DB, {Snow} B, {Fleck} B, {Stangalini} M, {Jafarzadeh} S.
\newblock 2020{\natexlab{b}}, {Reply to: Signatures of sunspot oscillations and
  the case for chromospheric resonances}.
\newblock \emph{Nature Astronomy}.
\newblock (\doi{10.1038/s41550-020-1158-4})

\bibitem[{Stangalini} et~al.(2018){Stangalini}, {Jafarzadeh}, {Ermolli},
  {Erd{\'e}lyi}, {Jess}, {Keys}, {Giorgi}, {Murabito}, {Berrilli}, and {Del
  Moro}]{2018ApJ...869..110S}
{Stangalini} M, {Jafarzadeh} S, {Ermolli} I, {Erd{\'e}lyi} R, {Jess} DB, {Keys}
  PH, {Giorgi} F, {Murabito} M, {Berrilli} F, {Del Moro} D.
\newblock 2018, {Propagating Spectropolarimetric Disturbances in a Large
  Sunspot}.
\newblock \emph{\apj}, 869\penalty0 (2):\penalty0 110.
\newblock (\doi{10.3847/1538-4357/aaec7b})

\bibitem[{Keys} et~al.(2019){Keys}, {Reid}, {Mathioudakis}, {Shelyag},
  {Henriques}, {Hewitt}, {Del Moro}, {Jafarzadeh}, {Jess}, and
  {Stangalini}]{2019MNRAS.488L..53K}
{Keys} PH, {Reid} A, {Mathioudakis} M, {Shelyag} S, {Henriques} VMJ, {Hewitt}
  RL, {Del Moro} D, {Jafarzadeh} S, {Jess} DB, {Stangalini} M.
\newblock 2019, {The magnetic properties of photospheric magnetic bright points
  with high-resolution spectropolarimetry}.
\newblock \emph{\mnras}, 488\penalty0 (1):\penalty0 L53--L58.
\newblock (\doi{10.1093/mnrasl/slz097})

\bibitem[{Houston} et~al.(2020){Houston}, {Jess}, {Keppens}, {Stangalini},
  {Keys}, {Grant}, {Jafarzadeh}, {McFetridge}, {Murabito}, {Ermolli}, and
  {Giorgi}]{2020ApJ...892...49H}
{Houston} SJ, {Jess} DB, {Keppens} R, {Stangalini} M, {Keys} PH, {Grant} SDT,
  {Jafarzadeh} S, {McFetridge} LM, {Murabito} M, {Ermolli} I, {Giorgi} F.
\newblock 2020, {Magnetohydrodynamic Nonlinearities in Sunspot Atmospheres:
  Chromospheric Detections of Intermediate Shocks}.
\newblock \emph{\apj}, 892\penalty0 (1):\penalty0 49.
\newblock (\doi{10.3847/1538-4357/ab7a90})

\bibitem[{Snow} and {Hillier}(2020)]{2020A&A...637A..97S}
{Snow} B, {Hillier} A.
\newblock 2020, {Mode conversion of two-fluid shocks in a partially-ionised,
  isothermal, stratified atmosphere}.
\newblock \emph{\aap}, 637:\penalty0 A97.
\newblock (\doi{10.1051/0004-6361/202037848})

\bibitem[{Woolston}(2019)]{2019Natur.572..143W}
{Woolston} C.
\newblock 2019, {Meeting the challenges of research across Africa}.
\newblock \emph{\nat}, 572\penalty0 (7767):\penalty0 143--145.
\newblock (\doi{10.1038/d41586-019-02311-2})

\bibitem[{Jess} et~al.(2012){Jess}, {De Moortel}, {Mathioudakis}, {Christian},
  {Reardon}, {Keys}, and {Keenan}]{2012ApJ...757..160J}
{Jess} DB, {De Moortel} I, {Mathioudakis} M, {Christian} DJ, {Reardon} KP,
  {Keys} PH, {Keenan} FP.
\newblock 2012, {The Source of 3 Minute Magnetoacoustic Oscillations in Coronal
  Fans}.
\newblock \emph{\apj}, 757\penalty0 (2):\penalty0 160.
\newblock (\doi{10.1088/0004-637X/757/2/160})

\bibitem[{Morgan} and {Druckm{\"u}ller}(2014)]{2014SoPh..289.2945M}
{Morgan} H, {Druckm{\"u}ller} M.
\newblock 2014, {Multi-Scale Gaussian Normalization for Solar Image
  Processing}.
\newblock \emph{\solphys}, 289\penalty0 (8):\penalty0 2945--2955.
\newblock (\doi{10.1007/s11207-014-0523-9})

\bibitem[{Fleck} et~al.(2020){Fleck}, {Carlsson}, {Khomenko}, {Rempel},
  {Steiner}, and {Vigeesh}]{Bernhard}
{Fleck} B, {Carlsson} M, {Khomenko} E, {Rempel} M, {Steiner} O, {Vigeesh} G.
\newblock 2021, {Acoustic-gravity wave propagation characteristics in 3D
  radiation hydrodynamic simulations of the solar atmosphere}.
\newblock \emph{\ptrsa}, 379, 20200170.
\newblock (\doi{10.1098/rsta.2020.0170})

\bibitem[{Keys} et~al.(2020){Keys}, {Steiner}, and {Vigeesh}]{Peter}
{Keys} PH, {Steiner} O, {Vigeesh} G.
\newblock 2021, {On the effect of oscillatory phenomena on Stokes inversion
  results}.
\newblock \emph{\ptrsa}, 379, 20200182.
\newblock (\doi{10.1098/rsta.2020.0182})

\bibitem[{Khomenko} et~al.(2020){Khomenko}, {Collados}, {Vitas}, and
  {Gonz{\'{a}}lez-Morales}]{Elena}
{Khomenko} E, {Collados} M, {Vitas} N, {Gonz{\'{a}}lez-Morales} PA.
\newblock 2021, {Influence of ambipolar and Hall effects on vorticity in 3D
  simulations of magneto-convection}.
\newblock \emph{\ptrsa}, 379, 20200176.
\newblock (\doi{10.1098/rsta.2020.0176})

\bibitem[{Eklund} et~al.(){Eklund}, {Wedemeyer}, {Snow}, {Jess}, {Jafarzadeh},
  {Grant}, {Carlsson}, and {Szydlarski}]{Henrik}
{Eklund} H, {Wedemeyer} S, {Snow} B, {Jess} DB, {Jafarzadeh} S, {Grant} SDT,
  {Carlsson} M, {Szydlarski} M.
\newblock 2021, {Characterisation of shock wave signatures at millimetre wavelengths from Bifrost simulations}.
\newblock \emph{\ptrsa}, 379, 20200185.
\newblock (\doi{10.1098/rsta.2020. 0185})

\bibitem[{Wootten} and {Thompson}(2009)]{2009IEEEP..97.1463W}
{Wootten} A, {Thompson} AR.
\newblock 2009, {The Atacama Large Millimeter/Submillimeter Array}.
\newblock \emph{IEEE Proceedings}, 97\penalty0 (8):\penalty0 1463--1471.
\newblock (\doi{10.1109/JPROC.2009.2020572})

\bibitem[{Wedemeyer} et~al.(2016){Wedemeyer}, {Bastian}, {Braj{\v{s}}a},
  {Hudson}, {Fleishman}, {Loukitcheva}, {Fleck}, {Kontar}, {De Pontieu},
  {Yagoubov}, {Tiwari}, {Soler}, {Black}, {Antolin}, {Scullion}, {Gun{\'a}r},
  {Labrosse}, {Ludwig}, {Benz}, {White}, {Hauschildt}, {Doyle}, {Nakariakov},
  {Ayres}, {Heinzel}, {Karlicky}, {Van Doorsselaere}, {Gary}, {Alissandrakis},
  {Nindos}, {Solanki}, {Rouppe van der Voort}, {Shimojo}, {Kato},
  {Zaqarashvili}, {Perez}, {Selhorst}, and {Barta}]{2016SSRv..200....1W}
{Wedemeyer} S, {Bastian} T, {Braj{\v{s}}a} R, {Hudson} H, {Fleishman} G,
  {Loukitcheva} M, {Fleck} B, {Kontar} EP, {De Pontieu} B, {Yagoubov} P,
  {Tiwari} SK, {Soler} R, {Black} JH, {Antolin} P, {Scullion} E, {Gun{\'a}r} S,
  {Labrosse} N, {Ludwig} HG, {Benz} AO, {White} SM, {Hauschildt} P, {Doyle} JG,
  {Nakariakov} VM, {Ayres} T, {Heinzel} P, {Karlicky} M, {Van Doorsselaere} T,
  {Gary} D, {Alissandrakis} CE, {Nindos} A, {Solanki} SK, {Rouppe van der
  Voort} L, {Shimojo} M, {Kato} Y, {Zaqarashvili} T, {Perez} E, {Selhorst} CL,
  {Barta} M.
\newblock 2016, {Solar Science with the Atacama Large Millimeter/Submillimeter
  Array{\textemdash}A New View of Our Sun}.
\newblock \emph{\ssr}, 200\penalty0 (1-4):\penalty0 1--73.
\newblock (\doi{10.1007/s11214-015-0229-9})

\bibitem[{Newington} and {Cally}(2010)]{2010MNRAS.402..386N}
{Newington} ME, {Cally} PS.
\newblock 2010, {Reflection and conversion of magnetogravity waves in the solar
  chromosphere: windows to the upper atmosphere}.
\newblock \emph{\mnras}, 402\penalty0 (1):\penalty0 386--394.
\newblock (\doi{10.1111/j.1365-2966.2009.15884.x})

\bibitem[{Kneer} and {Bello Gonz{\'a}lez}(2011)]{2011A&A...532A.111K}
{Kneer} F, {Bello Gonz{\'a}lez} N.
\newblock 2011, {On acoustic and gravity waves in the solar photosphere and
  their energy transport}.
\newblock \emph{\aap}, 532:\penalty0 A111.
\newblock (\doi{10.1051/0004-6361/201116537})

\bibitem[{Vigeesh} et~al.(2017){Vigeesh}, {Jackiewicz}, and
  {Steiner}]{2017ApJ...835..148V}
{Vigeesh} G, {Jackiewicz} J, {Steiner} O.
\newblock 2017, {Internal Gravity Waves in the Magnetized Solar Atmosphere. I.
  Magnetic Field Effects}.
\newblock \emph{\apj}, 835\penalty0 (2):\penalty0 148.
\newblock (\doi{10.3847/1538-4357/835/2/148})

\bibitem[{Vigeesh} and {Roth}(2020)]{2020A&A...633A.140V}
{Vigeesh} G, {Roth} M.
\newblock 2020, {Synthetic observations of internal gravity waves in the solar
  atmosphere}.
\newblock \emph{\aap}, 633:\penalty0 A140.
\newblock (\doi{10.1051/0004-6361/201936846})

\bibitem[{Straus} et~al.(2008){Straus}, {Fleck}, {Jefferies}, {Cauzzi},
  {McIntosh}, {Reardon}, {Severino}, and {Steffen}]{2008ApJ...681L.125S}
{Straus} T, {Fleck} B, {Jefferies} SM, {Cauzzi} G, {McIntosh} SW, {Reardon} K,
  {Severino} G, {Steffen} M.
\newblock 2008, {The Energy Flux of Internal Gravity Waves in the Lower Solar
  Atmosphere}.
\newblock \emph{\apjl}, 681\penalty0 (2):\penalty0 L125.
\newblock (\doi{10.1086/590495})

\bibitem[{Calchetti} et~al.(2020){Calchetti}, {Jefferies}, {Fleck}, {Berrilli},
  and {Shcherbik}]{Daniele}
{Calchetti} D, {Jefferies} SM, {Fleck} B, {Berrilli} F, {Shcherbik} DV.
\newblock 2021, {A New Method for Detecting Solar Atmospheric Gravity Waves}.
\newblock \emph{\ptrsa}, 379, 20200178.
\newblock (\doi{10.1098/rsta. 2020.0178})

\bibitem[{Rutten} and {Krijger}(2003)]{Rutten2003}
{Rutten} RJ, {Krijger} JM.
\newblock 2003, {Dynamics of the solar chromosphere IV. Evidence for
  atmospheric gravity waves from TRACE}.
\newblock \emph{\aap}, 407:\penalty0 735--740.
\newblock (\doi{10.1051/0004-6361:20030894})

\bibitem[{Freytag} et~al.(2012){Freytag}, {Steffen}, {Ludwig},
  {Wedemeyer-B{\"o}hm}, {Schaffenberger}, and {Steiner}]{cobold}
{Freytag} B, {Steffen} M, {Ludwig} HG, {Wedemeyer-B{\"o}hm} S, {Schaffenberger}
  W, {Steiner} O.
\newblock 2012, {Simulations of stellar convection with CO5BOLD}.
\newblock \emph{Journal of Computational Physics}, 231\penalty0 (3):\penalty0
  919--959.
\newblock (\doi{10.1016/j.jcp.2011.09.026})

\bibitem[{Vigeesh} et~al.(2020){Vigeesh}, {Roth}, {Steiner}, and
  {Fleck}]{Gangadharan}
{Vigeesh} G, {Roth} M, {Steiner} O, {Fleck} B.
\newblock 2021, {On the influence of magnetic topology on the propagation of
  internal gravity waves in the solar atmosphere}.
\newblock \emph{\ptrsa}, 379, 20200177.
\newblock (\doi{10.1098/rsta.2020.0177})

\bibitem[{Jafarzadeh} et~al.(2020){Jafarzadeh}, {Wedemeyer}, {Fleck},
  {Stangalini}, {Jess}, {Morton}, {Szydlarski}, {Henriques}, {Zhu},
  {Wiegelmann}, {Guevara G\'{o}mez}, {Grant}, {Chen}, {Reardon}, and
  {White}]{Shahin}
{Jafarzadeh} S, {Wedemeyer} S, {Fleck} B, {Stangalini} M, {Jess} DB, {Morton}
  RJ, {Szydlarski} M, {Henriques} VMJ, {Zhu} X, {Wiegelmann} T, {Guevara
  G\'{o}mez} JC, {Grant} SDT, {Chen} B, {Reardon} K, {White} SM.
\newblock 2021, {An overall view of temperature oscillations in the solar
  chromosphere with ALMA}.
\newblock \emph{\ptrsa}, 379, 20200174.
\newblock (\doi{10.1098/rsta.2020.0174})

\bibitem[{Guevara~G{\'{o}}mez} et~al.(2020){Guevara~G{\'{o}}mez}, {Jafarzadeh},
  {Wedemeyer}, {Szydlarski}, {Stangalini}, {Fleck}, and {Keys}]{Juan}
{Guevara~G{\'{o}}mez} JC, {Jafarzadeh} S, {Wedemeyer} S, {Szydlarski} M,
  {Stangalini} M, {Fleck} B, {Keys} PH.
\newblock 2021, {High-frequency oscillations in small chromospheric bright
  features observed with ALMA}.
\newblock \emph{\ptrsa}, 379, 20200184.
\newblock (\doi{10.1098/rsta.2020.0184})

\bibitem[{Gilchrist-Millar} et~al.(2020){Gilchrist-Millar}, {Jess}, {Grant},
  {Keys}, {Beck}, {Jafarzadeh}, {Riedl}, {Van Doorsselaere}, and
  {Cobo}]{Caitlin}
{Gilchrist-Millar} C, {Jess} DB, {Grant} S, {Keys} P, {Beck} C, {Jafarzadeh} S,
  {Riedl} JM, {Van Doorsselaere} T, {Cobo} BR.
\newblock 2021, {Magnetoacoustic Wave Energy Dissipation in the Atmosphere of
  Solar Pores}.
\newblock \emph{\ptrsa}, 379, 20200172.
\newblock (\doi{10.1098/rsta.2020.0172})

\bibitem[{Sych} et~al.(2020){Sych}, {Jess}, and {Su}]{Robert}
{Sych} R, {Jess} DB, {Su} J.
\newblock 2021, {The dynamics of 3-minute wavefronts and their relation to
  sunspot magnetic fields}.
\newblock \emph{\ptrsa}, 379, 20200180.
\newblock (\doi{10.1098/rsta.2020.0180})

\bibitem[{Albidah} et~al.(2020){Albidah}, {Brevis}, {Fedun}, {Ballai}, {Jess},
  {Stangalini}, {Higham}, and {Verth}]{Abdulrahman}
{Albidah} AB, {Brevis} W, {Fedun} V, {Ballai} I, {Jess} DB, {Stangalini} M,
  {Higham} J, {Verth} G.
\newblock 2021, {Proper Orthogonal and Dynamic Mode Decomposition of Sunspot
  Data}.
\newblock \emph{\ptrsa}, 379, 20200181.
\newblock (\doi{10.1098/rsta.2020.0181})

\bibitem[{Cavallini}(2006)]{2006SoPh..236..415C}
{Cavallini} F.
\newblock 2006, {IBIS: A New Post-Focus Instrument for Solar Imaging
  Spectroscopy}.
\newblock \emph{\solphys}, 236\penalty0 (2):\penalty0 415--439.
\newblock (\doi{10.1007/s11207-006-0103-8})

\bibitem[{Stangalini} et~al.(2020){Stangalini}, {Baker}, {Valori},
  {Jafarzadeh}, {Murabito}, {To}, {Brooks}, {Ermolli}, {Giorgi}, and
  {MacBride}]{Marco}
{Stangalini} M, {Baker} D, {Valori} JDB G., {Jafarzadeh} S, {Murabito} M, {To}
  ASH, {Brooks} DH, {Ermolli} I, {Giorgi} F, {MacBride} CD.
\newblock 2021, {Spectropolarimetric Fluctuations in a Sunspot Chromosphere}.
\newblock \emph{\ptrsa}, 379, 20200216.
\newblock (\doi{10.1098/rsta.2020.0216})

\bibitem[{Laming}(2015)]{2015LRSP...12....2L}
{Laming} JM.
\newblock 2015, {The FIP and Inverse FIP Effects in Solar and Stellar Coronae}.
\newblock \emph{Living Reviews in Solar Physics}, 12\penalty0 (1):\penalty0 2.
\newblock (\doi{10.1007/lrsp-2015-2})

\bibitem[{MacBride} et~al.(2020){MacBride}, {Jess}, {Grant}, {Khomenko},
  {Keys}, and {Stangalini}]{Conor}
{MacBride} CD, {Jess} DB, {Grant} SDT, {Khomenko} E, {Keys} PH, {Stangalini} M.
\newblock 2021, {Accurately constraining velocity information from spectral
  imaging observations using machine learning techniques}.
\newblock \emph{\ptrsa}, 379, 20200171.
\newblock (\doi{:10.1098/rsta.2020.0171})

\bibitem[{Morton} et~al.(2020){Morton}, {Mooroogen}, and {Henriques}]{Richard}
{Morton} RJ, {Mooroogen} K, {Henriques} VMJ.
\newblock 2021, {Transverse motions in sunspot super-penumbral fibrils}.
\newblock \emph{\ptrsa}, 379, 20200183.
\newblock (\doi{10.1098/rsta.2020.0183})

\bibitem[{Norton} et~al.(2020){Norton}, {Stutz}, and {Welsch}]{Aimee}
{Norton} AA, {Stutz} RB, {Welsch} BT.
\newblock 2021, {Oscillations observed in Umbra, Plage, Quiet-Sun and the
  Polarity Inversion Line of Active Region 11158 using HMI/SDO Data}.
\newblock \emph{\ptrsa}, 379, 20200175.
\newblock (\doi{10.1098/rsta.2020.0175})

\bibitem[{Scherrer} et~al.(2012){Scherrer}, {Schou}, {Bush}, {Kosovichev},
  {Bogart}, {Hoeksema}, {Liu}, {Duvall}, {Zhao}, {Title}, {Schrijver},
  {Tarbell}, and {Tomczyk}]{2012SoPh..275..207S}
{Scherrer} PH, {Schou} J, {Bush} RI, {Kosovichev} AG, {Bogart} RS, {Hoeksema}
  JT, {Liu} Y, {Duvall} TL, {Zhao} J, {Title} AM, {Schrijver} CJ, {Tarbell} TD,
  {Tomczyk} S.
\newblock 2012, {The Helioseismic and Magnetic Imager (HMI) Investigation for
  the Solar Dynamics Observatory (SDO)}.
\newblock \emph{\solphys}, 275\penalty0 (1-2):\penalty0 207--227.
\newblock (\doi{10.1007/s11207-011-9834-2})

\bibitem[{Hasan} et~al.(2010){Hasan}, {Soltau}, {K{\"a}rcher}, {S{\"u}{\ss}},
  and {Berkefeld}]{2010AN....331..628H}
{Hasan} SS, {Soltau} D, {K{\"a}rcher} H, {S{\"u}{\ss}} M, {Berkefeld} T.
\newblock 2010, {NLST: India's National Large Solar Telescope}.
\newblock \emph{Astronomische Nachrichten}, 331\penalty0 (6):\penalty0 628.
\newblock (\doi{10.1002/asna.201011389})

\bibitem[{Collados} et~al.(2010){Collados}, {Bettonvil}, {Cavaller}, {Ermolli},
  {Gelly}, {P{\'e}rez}, {Socas-Navarro}, {Soltau}, {Volkmer}, and {EST
  Team}]{2010AN....331..615C}
{Collados} M, {Bettonvil} F, {Cavaller} L, {Ermolli} I, {Gelly} B, {P{\'e}rez}
  A, {Socas-Navarro} H, {Soltau} D, {Volkmer} R, {EST Team}.
\newblock 2010, {European Solar Telescope: Progress status}.
\newblock \emph{Astronomische Nachrichten}, 331\penalty0 (6):\penalty0 615.
\newblock (\doi{10.1002/asna.201011386})

\bibitem[{Solanki} et~al.(2010){Solanki}, {Barthol}, {Danilovic}, {Feller},
  {Gandorfer}, {Hirzberger}, {Riethm{\"u}ller}, {Sch{\"u}ssler}, {Bonet},
  {Mart{\'\i}nez Pillet}, {del Toro Iniesta}, {Domingo}, {Palacios},
  {Kn{\"o}lker}, {Bello Gonz{\'a}lez}, {Berkefeld}, {Franz}, {Schmidt}, and
  {Title}]{2010ApJ...723L.127S}
{Solanki} SK, {Barthol} P, {Danilovic} S, {Feller} A, {Gandorfer} A,
  {Hirzberger} J, {Riethm{\"u}ller} TL, {Sch{\"u}ssler} M, {Bonet} JA,
  {Mart{\'\i}nez Pillet} V, {del Toro Iniesta} JC, {Domingo} V, {Palacios} J,
  {Kn{\"o}lker} M, {Bello Gonz{\'a}lez} N, {Berkefeld} T, {Franz} M, {Schmidt}
  W, {Title} AM.
\newblock 2010, {SUNRISE: Instrument, Mission, Data, and First Results}.
\newblock \emph{\apjl}, 723\penalty0 (2):\penalty0 L127--L133.
\newblock (\doi{10.1088/2041-8205/723/2/L127})

\bibitem[{Solanki} et~al.(2017){Solanki}, {Riethm{\"u}ller}, {Barthol},
  {Danilovic}, {Deutsch}, {Doerr}, {Feller}, {Gandorfer}, {Germerott}, {Gizon},
  {Grauf}, {Heerlein}, {Hirzberger}, {Kolleck}, {Lagg}, {Meller}, {Tomasch},
  {van Noort}, {Blanco Rodr{\'\i}guez}, {Gasent Blesa}, {Balaguer Jim{\'e}nez},
  {Del Toro Iniesta}, {L{\'o}pez Jim{\'e}nez}, {Orozco Suarez}, {Berkefeld},
  {Halbgewachs}, {Schmidt}, {{\'A}lvarez-Herrero}, {Sabau-Graziati}, {P{\'e}rez
  Grand e}, {Mart{\'\i}nez Pillet}, {Card}, {Centeno}, {Kn{\"o}lker}, and
  {Lecinski}]{2017ApJS..229....2S}
{Solanki} SK, {Riethm{\"u}ller} TL, {Barthol} P, {Danilovic} S, {Deutsch} W,
  {Doerr} HP, {Feller} A, {Gandorfer} A, {Germerott} D, {Gizon} L, {Grauf} B,
  {Heerlein} K, {Hirzberger} J, {Kolleck} M, {Lagg} A, {Meller} R, {Tomasch} G,
  {van Noort} M, {Blanco Rodr{\'\i}guez} J, {Gasent Blesa} JL, {Balaguer
  Jim{\'e}nez} M, {Del Toro Iniesta} JC, {L{\'o}pez Jim{\'e}nez} AC, {Orozco
  Suarez} D, {Berkefeld} T, {Halbgewachs} C, {Schmidt} W, {{\'A}lvarez-Herrero}
  A, {Sabau-Graziati} L, {P{\'e}rez Grand e} I, {Mart{\'\i}nez Pillet} V,
  {Card} G, {Centeno} R, {Kn{\"o}lker} M, {Lecinski} A.
\newblock 2017, {The Second Flight of the Sunrise Balloon-borne Solar
  Observatory: Overview of Instrument Updates, the Flight, the Data, and First
  Results}.
\newblock \emph{\apjs}, 229\penalty0 (1):\penalty0 2.
\newblock (\doi{10.3847/1538-4365/229/1/2})

\bibitem[{M{\"u}ller} et~al.(2020){M{\"u}ller}, {St. Cyr}, {Zouganelis},
  {Gilbert}, {Marsden}, {Nieves-Chinchilla}, {Antonucci}, {Auch{\`e}re},
  {Berghmans}, {Horbury}, {Howard}, {Krucker}, {Maksimovic}, {Owen}, {Rochus},
  {Rodriguez-Pacheco}, {Romoli}, {Solanki}, {Bruno}, {Carlsson}, {Fludra},
  {Harra}, {Hassler}, {Livi}, {Louarn}, {Peter}, {Sch{\"u}hle}, {Teriaca}, {del
  Toro Iniesta}, {Wimmer-Schweingruber}, {Marsch}, {Velli}, {De Groof},
  {Walsh}, and {Williams}]{2020A&A...642A...1M}
{M{\"u}ller} D, {St. Cyr} OC, {Zouganelis} I, {Gilbert} HR, {Marsden} R,
  {Nieves-Chinchilla} T, {Antonucci} E, {Auch{\`e}re} F, {Berghmans} D,
  {Horbury} TS, {Howard} RA, {Krucker} S, {Maksimovic} M, {Owen} CJ, {Rochus}
  P, {Rodriguez-Pacheco} J, {Romoli} M, {Solanki} SK, {Bruno} R, {Carlsson} M,
  {Fludra} A, {Harra} L, {Hassler} DM, {Livi} S, {Louarn} P, {Peter} H,
  {Sch{\"u}hle} U, {Teriaca} L, {del Toro Iniesta} JC, {Wimmer-Schweingruber}
  RF, {Marsch} E, {Velli} M, {De Groof} A, {Walsh} A, {Williams} D.
\newblock 2020, {The Solar Orbiter mission. Science overview}.
\newblock \emph{\aap}, 642:\penalty0 A1.
\newblock (\doi{10.1051/0004-6361/202038467})

\bibitem[{Shimizu} et~al.(2019){Shimizu}, {Imada}, {Kawate}, {Ichimoto},
  {Suematsu}, {Hara}, {Katsukawa}, {Kubo}, {Toriumi}, {Watanabe}, {Yokoyama},
  {Korendyke}, {Warren}, {Tarbell}, {De Pontieu}, {Teriaca}, {Sch{\"u}hle},
  {Solanki}, {Harra}, {Matthews}, {Fludra}, {Auch{\`e}re}, {Andretta},
  {Naletto}, and {Zhukov}]{2019SPIE11118E..07S}
{Shimizu} T, {Imada} S, {Kawate} T, {Ichimoto} K, {Suematsu} Y, {Hara} H,
  {Katsukawa} Y, {Kubo} M, {Toriumi} S, {Watanabe} T, {Yokoyama} T, {Korendyke}
  CM, {Warren} HP, {Tarbell} T, {De Pontieu} B, {Teriaca} L, {Sch{\"u}hle} UH,
  {Solanki} S, {Harra} LK, {Matthews} S, {Fludra} A, {Auch{\`e}re} F,
  {Andretta} V, {Naletto} G, {Zhukov} A.
\newblock 2019, {The Solar-C\_EUVST mission}.
\newblock In \emph{UV, X-Ray, and Gamma-Ray Space Instrumentation for Astronomy
  XXI}, 1111807volume 11118 of \emph{Society of Photo-Optical Instrumentation
  Engineers (SPIE) Conference Series}pages.
\newblock (\doi{10.1117/12.2528240})

\bibitem[{Jaeggli} et~al.(2019){Jaeggli}, {Anan}, {Kramar}, and
  {Lin}]{2019AAS...23410612J}
{Jaeggli} SA, {Anan} T, {Kramar} M, {Lin} H.
\newblock 2019, {Optical Alignment of DL-NIRSP Spectrograph}.
\newblock In \emph{American Astronomical Society Meeting Abstracts \#234},
  106.12volume 234 of \emph{American Astronomical Society Meeting
  Abstracts}pages

\end{thebibliography}

\end{document}